\documentclass[prd, superscriptaddress,nofootinbib]{revtex4}
\usepackage{graphicx}
\usepackage{dcolumn}
\usepackage{bm}

\def\r{\rho}

\begin{document}
%


\title{Remark on non-existence of self-similar solutions in $4+1$ vacuum gravitational collapse}

\author{Piotr Bizo\'n}
\affiliation{M. Smoluchowski Institute of Physics, Jagellonian University, Krak\'ow, Poland}
\author{Arthur Wasserman}
\affiliation{Department of Mathematics,
   University of Michigan, Ann Arbor, Michigan}
\date{\today}
\begin{abstract}
We give a very short proof that the vacuum Einstein equations in $4+1$ dimensions have no
cohomogeneity-two Bianchi IX continuously self-similar solutions.
\end{abstract}

\pacs{Valid PACS appear here}
\maketitle

\noindent \emph{Introduction.} In a recent paper \cite{bcs} it was shown that in five spacetime
dimensions one can perform a consistent cohomogeneity-two symmetry reduction of the vacuum Einstein
equations which -- in contrast to the spherically symmetric reduction -- admits time dependent
asymptotically flat solutions.  The key idea was to modify the standard spherically symmetric
ansatz by replacing the round metric on the three-sphere with the homogeneously squashed metric,
thereby breaking the $SO(4)$ isometry to $SO(3)\times U(1)$. In this way the squashing parameter
becomes a dynamical degree of freedom and  Birkhoff's theorem is evaded. This model (which we shall
refer to as the BCS model) provides a simple theoretical setting for studying the dynamics of
gravitational collapse in vacuum. Numerical simulations
   indicate that the spherically symmetric solutions, Minkowski and Schwarzschild,
   play the role of attractors in the evolution
  of  generic regular initial data  (small and large ones,
  respectively) and the transition between
these two  outcomes of evolution exhibits a discretely self-similar critical behavior \cite{bcs}.
In this respect the BCS model is very similar to the Einstein-massless scalar field system
\cite{ch1,ch2,ch}. However, there is one interesting difference between these two models which we
want to point out here. The difference is concerned with the existence of continuously self-similar
(CSS) solutions. In \cite{ch3} Christodoulou  proved that the Einstein-massless scalar field system
possesses CSS solutions. These solutions, suitably truncated, provide examples of naked
singularities developing from regular initial data (however, being unstable \cite{ch4}, they do not
contradict the weak cosmic censorship conjecture). We will show below that the BCS model has no CSS
solutions. This result indicates that
  the  CSS naked singularities  found by Christodoulou for the self-gravitating
  massless scalar field are, in a sense, matter generated (mathematically, they are due to the fact
  that only derivatives of the scalar field appear in the equations).
  \vskip 0.2cm
\noindent   \emph{The BCS ansatz and self-similarity.} After \cite{bcs} we parametrize the metric
as follows
\begin{equation}\label{metric}
    ds^2= - A e^{-2\delta} dt^2 + A^{-1} dr^2 + \frac{1}{4} r^2
     \left(e^{2B}(\sigma_1^2 + \sigma_2^2) + e^{-4B} \sigma_3^2\right)\,,
\end{equation}
where $A$, $\delta$, and $B$ are functions of $(t,r)$, and $\sigma_i$ are left invariant one-forms
on $SU(2)$ which in terms of the Euler angles take the form
\begin{equation}\label{sigma}
    \sigma_1=\cos{\psi}\;d\theta+\sin{\psi}\sin{\theta}\;d\phi,\quad
    \sigma_2=-\sin{\psi}\;d\theta+\cos{\psi}\sin{\theta}\;d\phi,\quad
     \sigma_3=d\psi + \cos{\theta}\; d\phi.
\end{equation}
Substituting this ansatz into the vacuum Einstein equations we get the following system of PDEs
\begin{eqnarray}\label{mom}
  \partial_r A &=& - \frac{2 A}{r} \!+\frac{2}{3r} \left(4 e^{-2B}- e^{-8B}\right)- 2
  r
 \left( e^{2\delta} A^{-1} (\partial_t B)^2 + A (\partial_r B)^2\right)\,,\\
    \partial_t A &=& - 4 r A\, (\partial_t B)\, (\partial_r B)\,,\\
 \partial_r \delta &=& - 2 r \left(e^{2\delta} A^{-2} (\partial_t B)^2 +
    (\partial_r B)^2\right)\,,\\
\partial_t \left(e^{\delta} A^{-1} r^3 \partial_t B \right) & =& \partial_r
\left(e^{-\delta} A r^3 \partial_r B\right) + \frac{4}{3} e^{-\delta} r
\left(e^{-2B}-e^{-8B}\right)\,.
\end{eqnarray}
These equations have the scaling symmetry $(t,r) \rightarrow (\lambda t, \lambda r)$ so it is
natural to look for continuously self-similar (CSS) solutions, that is solutions which are scale
invariant.  Such solutions depend on a single variable $\rho=r/t$ and then the system (3-6) reduces
to ordinary differential equations (where prime is $d/d\rho$ and $Z=e^{\delta} \rho/A$)
\begin{eqnarray}\label{a}
\r A' & = & - 2A +\frac{2}{3}\left(4 e^{-2B}- e^{-8B}\right) - 2\r^2 A (1+Z^2) B'^2\,,\\
 A'&=& -4 \r A B'^2\,,\\
\r Z' &=& Z+2Z(1-Z^2) \r^2 B'^2\,,\\
B''&=&\frac{(2Z^2-3)B'+2\r^2(1-Z^4)B'^3}{\r(1-Z^2)}+\frac{4}{3}\frac{e^{-2B}-e^{-8B}}{\r^2 A
(1-Z^2)}\,.
\end{eqnarray}
The combination of equations (7) and (8) yields the constraint
\begin{equation}\label{con}
    3A -3A (1-Z^2) \r^2 B'^2-4 e^{-2B} + e^{-8B}=0.
\end{equation}
We are interested in regular solutions, where 'regular' means twice continuously differentiable. At
the origin regular  solutions must satisfy  the following initial conditions
\begin{equation}\label{ic}
    B(\r)\sim b\r^2,\quad A(\r)\sim 1-4b^2\r^4,\quad Z(\r)\sim \r\,,
\end{equation}
where we used the remaining scaling freedom to set $Z'(0)=1$ for convenience. It follows from
(\ref{ic}) and Eq.(9) that if $Z<1$ then $Z(\r)\geq \r$, hence there is an $\r_0$ such that
$Z(\r_0)=1$. Geometrically, $\r_0$ corresponds to the similarity horizon (the light cone of the
singularity).

\vspace{0.2cm} \noindent \emph{Proof.} We will show that solutions starting from initial conditions
(\ref{ic}) cannot be regular at $\r_0$. Assume for contradiction that the solution
$(A(\r),Z(\r),B(\r))$ is regular on the closed interval $I=\{\r: 0\leq \r\leq \r_0\}$. First, note
that the function $A$ is positive on $I$ since from Eq.(8) we have
$A(\r)=\exp\left(-4\int\limits_0^{\r} s B'(s)^2 ds\right)$. Second, it follows from Eq.(10) that if
$B'(\r_1)=0$ for some $\r_1$ then $B''(\r_1)$ has the same sign as $B(\r_1)$. Thus, the function
$B(\r)$ is monotone on $I$ and $B'(\r)$ has the sign of $b$.  Next,
 let us define the function
\begin{equation}\label{h}
H=8 e^{-2B}-5 e^{-8B} - 3 \r A B' - 3 A\,.
\end{equation}
With the use of this function (which we found by an arduous trial and error),  the rest of the
proof amounts to a one-line exercise in elementary calculus
 The initial conditions
(\ref{ic}) imply that $H(\r)\sim 9 b \r^2$ near the origin.
 Differentating $H$ and using
the constraint (\ref{con}), we obtain
\begin{equation}\label{dh}
H'+\left(\frac{1}{\r(1-Z^2)}+3 B'\right) H=27 e^{-8B} B'\,,
\end{equation}
hence $H(\r)$ cannot have a zero for $\r<\r_0$ because if $H(\r)=0$ then $H'(\r)$ has the same sign
as $B'(\r)$ and therefore $b$. Similarly, $H(\r_0)$ cannot vanish because L'Hopital's rule gives
$H'(\r_0)=54 e^{-2B(\r_0)} B'(\r_0)$. However,  $H(\r_0)$ must vanish for regular solutions, as
follows immediately from (\ref{dh}). This contradiction ends the proof.

\end{document}